\documentclass[sigconf]{acmart}
\AtBeginDocument{%
  }

\usepackage{xspace}
\usepackage{soul}
\usepackage{caption}
\usepackage{listings}
\usepackage{enumitem}
\usepackage{tcolorbox}
\tcbuselibrary{breakable}
\tcbset{
  breakable,     %
}
\usepackage[normalem]{ulem}
\usepackage{multirow}
\usepackage{multicol}

\copyrightyear{2026}
\acmYear{2026}
\setcopyright{cc}
\setcctype{by}
\acmConference[WWW Companion '26]{Companion Proceedings of the ACM Web Conference 2026}{April 13--17, 2026}{Dubai, United Arab Emirates}
\acmBooktitle{Companion Proceedings of the ACM Web Conference 2026 (WWW Companion '26), April 13--17, 2026, Dubai, United Arab Emirates}
\acmPrice{}
\acmDOI{10.1145/3774905.3793109}
\acmISBN{979-8-4007-2308-7/2026/04}

\begin{document}

\title[\autoresearcher: Transparent Multi-Agent Research Ideation]
      {\autoresearcher: Automating Knowledge-Grounded and Transparent Research Ideation with Multi-Agent Collaboration}

\author{Jiawei Zhou}
\affiliation{
  \institution{ACEM, Shanghai Jiao Tong University}
  \city{Shanghai}
  \country{China}
}
\email{davidzjw@sjtu.edu.cn}

\author{Ruicheng Zhu}
\affiliation{
  \institution{ACEM, Shanghai Jiao Tong University}
    \city{Shanghai}
  \country{China}
}
\email{zhuruicheng@sjtu.edu.cn}

\author{Mengshi Chen}
\affiliation{
  \institution{ACEM, Shanghai Jiao Tong University}
    \city{Shanghai}
  \country{China}
}
\email{chenmengshi@sjtu.edu.cn}

\author{Jianwei Wang}
\authornote{Jianwei Wang and Kai Wang are joint corresponding authors.}
\affiliation{
  \institution{University of New South Wales}
    \city{Sydney}
  \country{Australia}
}
\email{jianwei.wang1@unsw.edu.au}

\author{Kai Wang}
\authornotemark[1]
\affiliation{
  \institution{ACEM, Shanghai Jiao Tong University}
  \city{Shanghai}
  \country{China}
}
\email{w.kai@sjtu.edu.cn}

\renewcommand{\shortauthors}{Zhou et al.}

\newcommand{\kwnospace}[1]{{\textsc{#1}}}
\newcommand{\autoresearcher}{\kwnospace{TrustResearcher}\xspace}
\newcommand{\AutoResearcher}{\autoresearcher}
\begin{abstract}

Agentic systems have recently emerged as a promising tool to automate literature-based ideation. However, current systems often remain black-box, with limited transparency or control for researchers.
Our work introduces \autoresearcher, a multi-agent demo system for knowledge-grounded and transparent ideation. 
Specifically, \autoresearcher integrates meticulously designed four stages into a unified framework: (A) Structured Knowledge Curation, (B) Diversified Idea Generation, (C) Multi-stage Idea Selection, and (D) Expert Panel Review \& Synthesis. 
Different from prior pipelines, our system not only exposes intermediate reasoning states, execution logs, and configurable agents for inspections, but also enables diverse and evidence-aligned idea generation. 
Our design is also domain-agnostic, where the same pipeline can be instantiated in any scientific field. 
As an illustrative case, we demonstrate \autoresearcher on a graph-mining scenario ($k$-truss breaking problem), where it generates distinct, plausible candidates with evidence and critiques. A live demo and source code are available at \url{https://github.com/valleysprings/TrustResearcher}.
\end{abstract}




\maketitle

\section{Introduction}
\label{sec:intro}

The formulation of novel research ideas is a central driver of scientific progress. It remains one of the most challenging and time-intensive stages of research-oriented inquiries, since effective research depends on organizing large volumes of information and cultivating original, diverse, and innovative solutions. As many fields expand at an unprecedented pace, researchers confront severe information overload: the literature has grown beyond what any individual can reasonably process. At the same time, cognitive constraints, such as bias, fixation, and narrow search strategies, further limit the exploration of genuinely novel and primitive ideas. This tension between accelerating knowledge production and bounded human attention creates a bottleneck for innovation.

Foundation models, particularly large language models (LLMs), can rapidly reorganize knowledge at scales beyond human capacity ~\cite{,zhao2023survey}. 
Emerging evidence suggests that they can foster creative and divergent thinking across scientific domains~\cite{baek2025researchagent,radensky2025scideator,Si2025Can,yamada2025aiscientist}.

Prior work motivating our design follows two related but distinct directions. 
First, LLM-based \emph{ideation scaffolds} investigate how interaction patterns and prompting strategies expand candidate spaces and alleviate cognitive fixation, without necessarily enforcing persistent evidence alignment across iterations~\cite{liu2024breadthdepth}. 
Second, \emph{agentic, grounded ideation pipelines} treat retrieval and structured knowledge (e.g., KGs) as integral to the workflow, coupling them with planning and critique to support traceable and verifiable research proposals~\cite{baek2025researchagent,Si2025Can,li2024chainofideas,yamada2025aiscientist}.

However, existing systems rarely achieve this balance, leaving automated research ideation bottlenecked by two fundamental gaps.
First, the absence of multi-stage, granular grounding. Existing systems typically treat retrieval as a simple and monolithic stage (especially for some current benchmark~\cite{guo2024ideabench}), failing to maintain overall comprehension of ideas throughout the iterative reasoning process. This results in a trade-off where open-ended prompts trigger ungrounded hallucinations, while rigid constraints stifle the discovery of primitive and original ideas~\cite{baek2025researchagent,liu2024breadthdepth,radensky2025scideator}. Second, the opacity of agentic coordination. Current multi-agent workflows often function as black boxes, lacking the operational transparency and auditable execution logs necessary for researchers to inspect internal logic or verify reasoning trajectories \cite{yamada2025aiscientist}. Such limitations have been widely recognized as a core challenge for agentic research \cite{wei2025agenticscience}.

To address these challenges, we introduce \autoresearcher, a multi-agent framework for knowledge-grounded and transparent research ideation.
Our work advances research ideation along two dimensions: system-level design and agent-level designs.

At the system level, we propose a compact four-stage design that mirrors the real-world use of ideation tools, improving efficiency and maintaining methodological control.
Specifically, it consists of
\textbf{(A) Structured Knowledge Curation}: Anchors the process through topic decomposition, retrieval, and KG construction, organizing evidence into a structured and traceable context;
\textbf{(B) Diversified Idea Generation}: Transforms the grounded context into diverse yet structured idea candidates via planning, decomposition, and diverse idea generation strategies;
\textbf{(C) Multi-stage Idea Selection}: Combines internal scoring with external similarity checks against the retrieved literature to filter redundant or weakly-supported candidates;
\textbf{(D) Expert Panel Review \& Synthesis}: Integrates parallel peer-style reviews to consolidate selected ideas into a coherent proposal. By linking planning, idea generation, idea selection, and critique to curated structured evidence, \autoresearcher promotes hypotheses that are traceable to prior works and relevant background knowledge.

At the agent level, \autoresearcher is realized through a sophisticated multi-agent orchestration that delegates specialized cognitive tasks to coordinated agents. The system’s intelligence emerges from the synergy between curation agents, which build a grounded conceptual backbone through multi-granularity retrieval and incremental KG construction, and planner agents that distill this knowledge into research blueprints. These blueprints guide specialized agents to explore reasoning trajectories, generating a diverse yet evidence-aligned candidate pool. To enforce rigor, an asynchronous expert panel conducts parallel, multi-dimensional critiques with filtering based on technical soundness and original contribution. This entire lifecycle is governed by a transparent orchestrator that manages iterative refinement while exposing the system’s underlying reasoning traces through auditable execution logs. \autoresearcher provides the robustness and transparency required to liberate researchers from cognitive overhead.

In summary, this work makes three key contributions:
\begin{itemize}
    \item We present \autoresearcher, a multi-agent system for knowledge-grounded and transparent research ideation.
    \item \autoresearcher comprises four components (A--D) system-wise. Agentic-wise, \autoresearcher features multi-stage / multi-granular knowledge grounding (task-decomposed paper retrieval with KG-based grounding), diversified idea generation with iterative self-refinement, orchestrated filtering and reviewing strategies and transparent intermediate artifacts with auditable execution traces.
    \item We release a web demonstration showcasing the interactive workflow and outputs of \autoresearcher on a $k$-truss-based graph mining task, illustrating how \autoresearcher can support real-world scientific research ideation.
\end{itemize}

\section{System Design}

\begin{figure*}[ht]
    \centering
    \includegraphics[width=\linewidth]{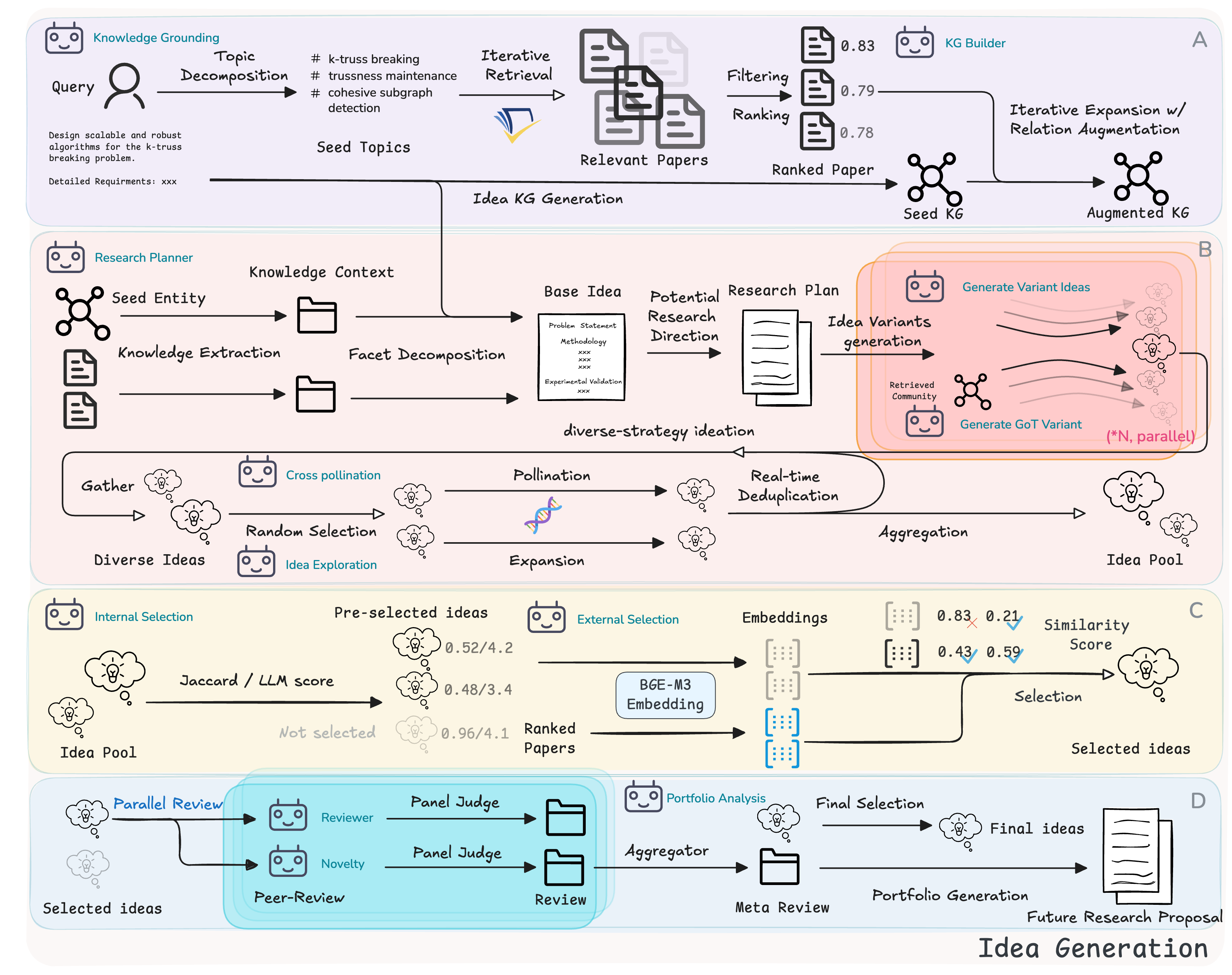}
    \caption{System architecture of \autoresearcher, illustrated with the $k$-truss breaking problem. The system comprises four modules connected in an end-to-end pipeline. Arrows denote steps handled by agents.}
    \label{fig:framework}
\end{figure*}

Figure~\ref{fig:framework} presents the architecture of \autoresearcher, which integrates four core modules: structured knowledge curation, diversified idea generation, idea selection, and expert panel review. Each module mirrors a corresponding phase in human research iteration, from active grounding in prior research to generating, refining, and evaluating hypotheses. Details of each component are provided in Sections~\ref{subsec:Knowledge}–\ref{subsec:Panel}.

\vspace{-2mm}
\subsection{Structured Knowledge Curation \label{subsec:Knowledge}}

Like human researchers who first survey existing literature before generating new ideas, \autoresearcher begins by constructing a structured KG that organizes retrieved papers into a traceable state. This KG provides well-grounded contexts for downstream idea generation. It comprises multi-granularity retrieval and incremental KG construction.

\noindent\textbf{Multi-granularity Retrieval}.
Given a seed topic, \autoresearcher first performs LLM-guided topic decomposition to identify salient domain concepts with varying granularity. The LLM then organizes these concepts into a set of well-formed search queries, rather than relying on a single handcrafted query. Each query is executed via the Semantic Scholar API under a fixed retrieval budget. The resulting papers are merged using semantic filtering and deduplication, with further pruning based on topic name overlap, to produce the final collection of paper samples. This multi-granularity retrieval process broadens coverage across related subtopics while maintaining topical relevance.

\noindent\textbf{Incremental KG Construction}.
To achieve balanced coverage and interpretability, \autoresearcher constructs the KG in four controlled phases. First, LLM-based entity extraction identifies core problems, methods, and applications to form the conceptual backbone. Second, mini-batch enrichment adds entities and relations from paper metadata while maintaining contextual coherence. Third, degree-based expansion samples top-$K$ high-degree nodes (default $K{=}10$) to mine adjacent or emerging concepts. Finally, hybrid sampling (60\% high-degree, 40\% random) uncovers latent methodological and theoretical links without introducing new entities. This incremental process yields a structured and extensible KG that combines precision with exploratory breadth.

\subsection{Diversified Idea Generation \label{subsec:GEN}}

Building on the curated KG, \autoresearcher generates research ideas through a hierarchical, multi-strategy process that mirrors human brainstorming, consisting of literature-informed planning, graph-of-thought exploration, multi-strategy variant generation, and iterative refinement. The goal is to produce a diverse yet coherent idea pool.

\noindent\textbf{Literature-Informed Planning}.
We introduce a planner agent that serves as the bridge between knowledge grounding and idea synthesis. It analyzes high-degree entities and summarizes semantically relevant papers to extract key research cues. Through LLM-based gap analysis, it identifies open limitations and decomposes them into three facets: \emph{Problem Statement}, \emph{Proposed Methodology} and \emph{Experimental Validation}. This structured blueprint provides the foundation for large-scale idea generation.

\noindent\textbf{Graph-of-Thought Exploration}.
To leverage structural grounding, we adopt \emph{Graph-of-Thought (GoT)} reasoning~\cite{besta2024graph} to sample KG-grounded reasoning traces. The GoT module connects facet nodes to $20$ high-degree KG entities and performs asynchronous depth-first sampling (branching factor $b{=}3$, depth $d{=}5$), scoring paths by node quality ($0.6$), edge-type diversity ($0.2$), and length preference ($0.2$). High-quality paths are retained, each encoding a distinct research trajectory that enriches idea diversity.

\noindent\textbf{Multi-Strategy Variant Generation}.
To ensure breadth, \autoresearcher applies an over-generation factor ($\alpha{=}10$) and executes $3$ strategies in parallel: (1) Base variants extend the faceted plan; (2) GoT variants reformulate high-scoring reasoning paths into structured proposals; and (3) Cross-pollination, triggered on demand, synthesizes hybrids of top-ranked ideas guided by KG cross-connections. Redundant ideas are pruned via real-time string and semantic matching.

\noindent\textbf{Iterative Refinement}.
The refined pool undergoes multi-round critique and validation elaboration, expanding experimental facets (datasets, metrics, performance evaluation, ablation) while preserving revision history, scores, and reasoning traces. The resulting transparent idea pool captures reasoning traces, revision history, and supporting literature.

\subsection{Idea Selection \label{subsec:Selection}}

To ensure both coherence and novelty, \autoresearcher performs a two-stage selection.

\noindent\textbf{Internal Selection}. Each idea is evaluated across four weighted criteria: novelty ($0.30$), feasibility ($0.25$), clarity ($0.20$), and impact ($0.25$) to generate preliminary rankings. Top ideas undergo iterative merging: pairs with Jaccard similarity $>0.85$ are LLM-merged into unified proposals until convergence criteria are met. This process eliminates redundancy via conceptual-level consolidation.

\noindent\textbf{External Selection}. Remaining ideas are compared to retrieved literature using BGE-M3 embeddings~\cite{chen2024m3} and selected using cosine similarity over combined text fields. Candidates with maximum similarity below $0.7$ are retained, while the system logs top overlapping papers for transparency.

\subsection{Expert Panel Review \& Synthesis \label{subsec:Panel}}
Inspired by scientific peer review, \autoresearcher employs a multi-agent evaluator to ensure that generated ideas are not only diverse and novel but also feasible, rigorous, and well-justified. Two specialized agents operate \emph{asynchronously} to review all selected ideas in parallel. The \emph{reviewer agent} focuses on technical soundness and feasibility, while the \emph{novelty agent} assesses originality and contribution relative to prior work. Both agents follow a structured scoring template derived from standard conference and journal review rubrics, evaluating each idea across five dimensions (feasibility, expected impact, technical soundness, implementation complexity, and distinctiveness from prior work) on a 1–5 scale, with qualitative feedback on strengths, weaknesses, and recommended revisions.

An aggregator module fuses both perspectives by averaging dimension-level scores (equal weighting) and consolidating textual feedback into a single meta-review. A unified score is then computed using a weighted aggregate that emphasizes feasibility and originality, with all sub-scores preserved for traceability. Ideas with unified scores above 3.5 (approximately corresponding to a “weak accept” in peer-review terms) are classified as \emph{high-quality} and prioritized for inclusion. If the number of high-quality ideas exceeds the target portfolio size, all are retained to avoid discarding promising work; if fewer meet the threshold, remaining slots are filled by the generated best candidates.

\section{Demonstration}\label{ssec:results}
We illustrate the user interface and a live case study of \AutoResearcher. Additional analyses are in the project repository.

\begin{figure*}[t]
    \centering
    \includegraphics[width=\linewidth]{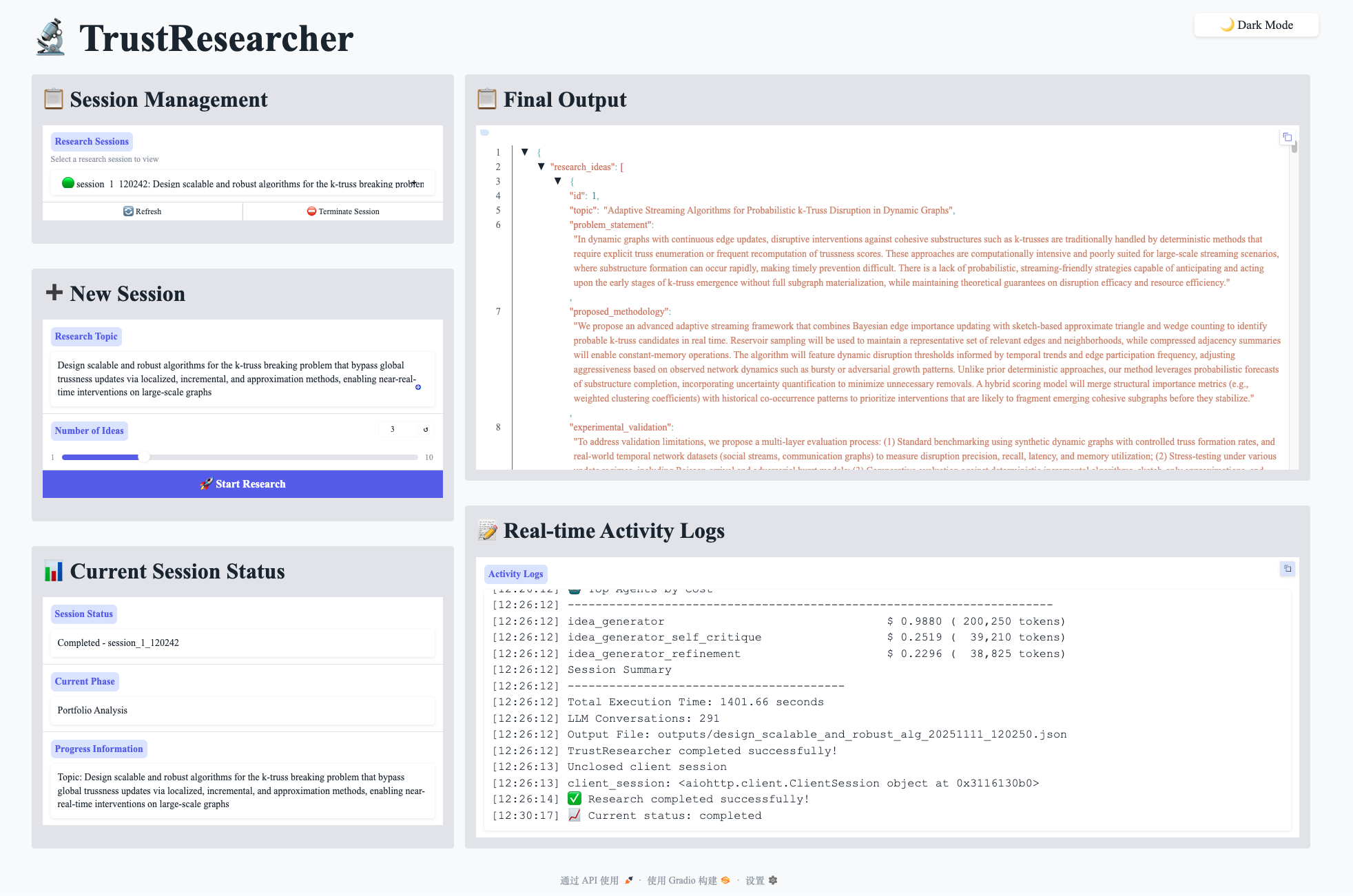}
    \caption{System interface during a live demonstration on the $k$-truss breaking problem.}
    \label{fig:visual}
\end{figure*}

\subsection{Interface}
Figure~\ref{fig:visual} presents the demonstration interface of \AutoResearcher. The interface is organized into four functional regions: (a) \emph{Session control}, where users define the research topic, number of ideas, and launch new sessions; (b) \emph{Status monitor}, displaying the current phase, runtime progress, and metadata; (c) \emph{Final output}, summarizing session results with structured downloads; and (d) \emph{Real-time logs}, showing detailed traces such as phase transitions, run-time, and agent activities. This layout emphasizes transparency and interactivity, where users can track the entire pipeline in real time, with each component’s output preserved as JSON files for inspection.

\subsection{Case study}

We demonstrate \AutoResearcher via a live case study on the \emph{$k$-truss breaking problem}~\cite{zhu2025truss}, a graph-mining primitive closely related to community search~\cite{fang2020survey}. In particular, $k$-truss structure is widely used to model cohesive subgraphs and is frequently adopted as a building block in community search pipelines; for example, \citet{zhou2025comet} studies community search with different cohesive structures or PPR pruning as structural priors to identify query-relevant communities, where truss-like cohesiveness constraints are often used to enforce structural quality.

Formally, given an undirected graph $G$ and an integer $k$, the $k$-truss breaking problem asks for a minimum-size edge set whose removal destroys all $k$-trusses in $G$. This problem is combinatorially challenging: removing a single edge may trigger cascading structural changes, and the problem remains under active investigation in the graph mining community. These characteristics make it a suitable stress test for \AutoResearcher’s ability to retrieve relevant literature, reason over incomplete knowledge, and explore novel research directions. 

This case study focuses on the research topic.

\begin{tcolorbox}[breakable, colback=green!5!white, colframe=green!70!black, title=Topic selected]
\emph{“Design scalable and robust algorithms for the \emph{k}-truss breaking problem that bypass global trussness updates via localized, incremental, and approximation methods, enabling near-real-time interventions on large-scale graphs.”} 

\end{tcolorbox}

The \autoresearcher system is prompted with this topic to generate multiple research hypotheses. Using the GPT-5 model, it produced three candidate research ideas within approximately 15-30 minutes (depending on the external LLM services), consuming over 200K tokens in total. Among these, we sample a technically coherent and novel one for in-depth analysis, presented below.

\begin{tcolorbox}[breakable, colback=orange!5!white, colframe=orange!70!black, title=\textbf{Scalable Approximation Framework for Real-Time \emph{k}-Truss Disruption in Massive Dynamic Graphs}]
\textbf{Motivation and Problem Statement.}
Modern large-scale and dynamically evolving graphs—such as social media interaction networks and communication infrastructures—often contain dense \emph{k}-truss substructures representing cohesive communities.
The \emph{k}-truss disruption problem seeks a cost-effective set of edge deletions that dismantle all \emph{k}-trusses.
Existing algorithms become computationally intractable as graph sizes reach billions of edges and undergo rapid updates, making real-time recomputation infeasible.
\textbf{There exists a critical methodological gap}: scalable, adaptive, and provably approximate frameworks for streaming, distributed data.

\vspace{0.6em}
\textbf{Novelty and Contributions.}
We propose a \emph{scalable approximation framework} integrating streaming graph analytics with probabilistic estimation and distributed computing principles.
The originality lies in its ability to perform real-time, approximation-guaranteed \emph{k}-truss disruption without recomputation. Key components include:
\begin{itemize}[leftmargin=1.8em, topsep=0.3em, itemsep=0.3em]
    \item \textbf{Incremental \emph{k}-truss maintenance} via sketch-based triangle estimation, leveraging bounded-error probabilistic counting.
    \item \textbf{Hierarchical graph partitioning} for computation locality and parallel updates.
    \item \textbf{Locality-sensitive edge prioritization} with adaptive scoring of disruption potential.
    \item \textbf{Adaptive deletion budgeting} aligned with temporal streaming windows.
    \item \textbf{Distributed integration} with Apache Flink / Spark Structured Streaming for near-linear scalability.
\end{itemize}

Approximation bounds are derived under probabilistic graph models (Chung–Lu, SBM), providing theoretical guarantees and tunable accuracy–speed trade-offs.

\vspace{0.6em}
\textbf{Experimental Validation.}
\begin{itemize}[leftmargin=1.8em, topsep=0.3em, itemsep=0.3em]
    \item \textbf{Scalability tests} on R-MAT, Barabási–Albert, and SBM graphs up to $10^9$ edges.
    \item \textbf{Streaming experiments} on Twitter and citation networks to assess disruption efficiency.
    \item \textbf{Benchmarking} against static and dynamic baselines.
    \item \textbf{Robustness analysis} under adversarial noise and delayed updates.
    \item \textbf{Ablation studies} isolating incremental and partitioning components.
    \item \textbf{Distributed stress tests} evaluating scalability and communication cost.
    \item \textbf{Case studies} on real-world crisis networks illustrating societal impact.
\end{itemize}

\vspace{0.6em}
\textbf{Summary.}
This framework provides the first \emph{provably efficient, real-time, distributed approximation method} for \emph{k}-truss disruption—bridging the gap between theoretical graph mining and actionable, scalable network intervention.
\end{tcolorbox}

Starting from the task description, \AutoResearcher retrieves and organizes relevant work on $k$-truss decomposition and related topics, including scalable decomposition algorithms, parallel index construction, and dynamic maintenance techniques. The retrieved literature is consolidated into a lightweight KG that serves as an explicit evidence base for subsequent reasoning.

Grounded in this curated context, the system explores multiple distinct research directions rather than optimizing a single solution. For the $k$-truss breaking task, the generated candidates span (i) localized scalable algorithms that avoid global recomputation, (ii) epidemic-containment–inspired strategies on temporal contact networks, and (iii) learning-based edge importance prediction. Each direction is instantiated as a structured proposal comprising a problem formulation, a methodological sketch, and an evaluation plan.

The candidate directions are progressively refined through similarity-based filtering and literature alignment checks to remove redundant or weak variants. Reviewer-style agents then assess the remaining proposals with numerical scores and qualitative critiques. For example, the localized algorithmic approach received high ratings for novelty (4.2) and clarity (4.5), while scalability on extremely large graphs was identified as a potential limitation. Aggregating these evaluations yields a final portfolio of five high-quality research candidates (average score $\approx$4.1/5), covering algorithmic, theoretical, distributed, and learning-based perspectives.

\vspace{-2mm}
\section{Conclusion}

In this work, we present \AutoResearcher, a multi-agent demo system for knowledge-grounded and transparent ideation built on four stages: Structured Knowledge Curation, Diversified Idea Generation, Multi-stage Idea Selection, and Expert Panel Review \& Synthesis.
It generates a broad, non-redundant set of candidate ideas, expanding the researcher’s space for exploration. Iterative self-refinement and knowledge grounding further ensure these ideas are technically sound and actionable. %

\bibliographystyle{ACM-Reference-Format}
\bibliography{sample}

@misc{radensky2025scideator,
  author = {Radensky, Marissa and Shahid, Simra and Fok, Raymond and Siangliulue, Pao and Hope, Tom and Weld, Daniel S.},
  title = {{Scideator}: Human-{LLM} Scientific Idea Generation Grounded in Research-Paper Facet Recombination},
  year = {2025},
  note = {Preprint, arXiv:2409.14634}
}

@misc{yamada2025aiscientist,
  author = {Yamada, Yutaro and Lange, Robert Tjarko and Lu, Cong and Hu, Shengran and Lu, Chris and Foerster, Jakob and Clune, Jeff and Ha, David},
  title = {The {AI} Scientist-v2: Workshop-Level Automated Scientific Discovery via Agentic Tree Search},
  year = {2025},
  note = {Preprint, arXiv:2504.08066}
}

@inproceedings{liu2024breadthdepth,
author = {Liu, Yiren and Chen, Si and Cheng, Haocong and Yu, Mengxia and Ran, Xiao and Mo, Andrew and Tang, Yiliu and Huang, Yun},
title = {How AI Processing Delays Foster Creativity: Exploring Research Question Co-Creation with an LLM-based Agent},
year = {2024},
isbn = {9798400703300},
publisher = {Association for Computing Machinery},
address = {New York, NY, USA},
url = {https://doi.org/10.1145/3613904.3642698},
doi = {10.1145/3613904.3642698},
booktitle = {Proceedings of the 2024 CHI Conference on Human Factors in Computing Systems},
articleno = {17},
numpages = {25},
location = {Honolulu, HI, USA},
series = {CHI '24}
}

@inproceedings{guo2024ideabench,
author = {Guo, Sikun and Shariatmadari, Amir Hassan and Xiong, Guangzhi and Huang, Albert and Kim, Myles and Williams, Corey M. and Bekiranov, Stefan and Zhang, Aidong},
title = {IdeaBench: Benchmarking Large Language Models for Research Idea Generation},
year = {2025},
isbn = {9798400714542},
publisher = {Association for Computing Machinery},
address = {New York, NY, USA},
url = {https://doi.org/10.1145/3711896.3737419},
doi = {10.1145/3711896.3737419},
booktitle = {Proceedings of the 31st ACM SIGKDD Conference on Knowledge Discovery and Data Mining V.2},
pages = {5888–5899},
numpages = {12},
location = {Toronto ON, Canada},
series = {KDD '25}
}

@inproceedings{baek2025researchagent,
    title = "{R}esearch{A}gent: Iterative Research Idea Generation over Scientific Literature with Large Language Models",
    author = "Baek, Jinheon  and
      Jauhar, Sujay Kumar  and
      Cucerzan, Silviu  and
      Hwang, Sung Ju",
    editor = "Chiruzzo, Luis  and
      Ritter, Alan  and
      Wang, Lu",
    booktitle = "Proceedings of the 2025 Conference of the Nations of the Americas Chapter of the Association for Computational Linguistics: Human Language Technologies (Volume 1: Long Papers)",
    month = apr,
    year = "2025",
    address = "Albuquerque, New Mexico",
    publisher = "Association for Computational Linguistics",
    url = "https://aclanthology.org/2025.naacl-long.342/",
    doi = "10.18653/v1/2025.naacl-long.342",
    pages = "6709--6738",
    ISBN = "979-8-89176-189-6"
}

@article{wei2025agenticscience,
  title        = {From AI for Science to Agentic Science: A Survey on Autonomous Scientific Discovery},
  author       = {Wei, Jiaqi and Yang, Yuejin and Zhang, Xiang and Chen, Yuhan and Zhuang, Xiang and Gao, Zhangyang and Zhou, Dongzhan and Wang, Guangshuai and Gao, Zhiqiang and Cao, Juntai and Qiu, Zijie and He, Xuming and Zhang, Qiang and You, Chenyu and Zheng, Shuangjia and Ding, Ning and Ouyang, Wanli and Dong, Nanqing and Cheng, Yu and Sun, Siqi and Bai, Lei and Zhou, Bowen},
  journal      = {arXiv preprint arXiv:2508.14111},
  year         = {2025},
  doi          = {10.48550/arXiv.2508.14111},
  url          = {https://arxiv.org/abs/2508.14111}
}

@inproceedings{li2024chainofideas,
    title = "Chain of Ideas: Revolutionizing Research Via Novel Idea Development with {LLM} Agents",
    author = "Li, Long  and
      Xu, Weiwen  and
      Guo, Jiayan  and
      Zhao, Ruochen  and
      Li, Xingxuan  and
      Yuan, Yuqian  and
      Zhang, Boqiang  and
      Jiang, Yuming  and
      Xin, Yifei  and
      Dang, Ronghao  and
      Rong, Yu  and
      Zhao, Deli  and
      Feng, Tian  and
      Bing, Lidong",
    editor = "Christodoulopoulos, Christos  and
      Chakraborty, Tanmoy  and
      Rose, Carolyn  and
      Peng, Violet",
    booktitle = "Findings of the Association for Computational Linguistics: EMNLP 2025",
    month = nov,
    year = "2025",
    address = "Suzhou, China",
    publisher = "Association for Computational Linguistics",
    url = "https://aclanthology.org/2025.findings-emnlp.477/",
    doi = "10.18653/v1/2025.findings-emnlp.477",
    pages = "8971--9004",
    ISBN = "979-8-89176-335-7"
}

@inproceedings{chen2024m3,
  title={M3-embedding: Multi-linguality, multi-functionality, multi-granularity text embeddings through self-knowledge distillation},
  author={Chen, Jianlyu and Xiao, Shitao and Zhang, Peitian and Luo, Kun and Lian, Defu and Liu, Zheng},
  booktitle={Findings of the Association for Computational Linguistics ACL 2024},
  pages={2318--2335},
  year={2024}
}

@inproceedings{besta2024graph,
author = {Besta, Maciej and Blach, Nils and Kubicek, Ales and Gerstenberger, Robert and Podstawski, Micha\l{} and Gianinazzi, Lukas and Gajda, Joanna and Lehmann, Tomasz and Niewiadomski, Hubert and Nyczyk, Piotr and Hoefler, Torsten},
title = {Graph of thoughts: solving elaborate problems with large language models},
year = {2024},
isbn = {978-1-57735-887-9},
publisher = {AAAI Press},
url = {https://doi.org/10.1609/aaai.v38i16.29720},
doi = {10.1609/aaai.v38i16.29720},
booktitle = {Proceedings of the Thirty-Eighth AAAI Conference on Artificial Intelligence},
articleno = {1972},
numpages = {9},
series = {AAAI'24}
}

@inproceedings{Si2025Can,
  title={{Can LLMs Generate Novel Research Ideas? A Large-Scale Human Study with 100+ NLP Researchers}},
  author={Chenglei Si and Diyi Yang and Tatsunori Hashimoto},
  year={2025},
  booktitle={ICLR},
  url={https://arxiv.org/abs/2409.04109}
}

@inproceedings{zhu2025truss,
  title = {Efficient K-{{Truss Breaking}} and {{Minimization}}},
  booktitle = {2025 {{IEEE}} 41st {{International Conference}} on {{Data Engineering}} ({{ICDE}})},
  author = {Zhu, Ruicheng and Wang, Xintong and Wang, Kai and Zhang, Fan and Qian, Zhengping and Yuan, Long},
  year = 2025,
  month = may,
  pages = {2628--2641},
  address = {Hong Kong},
  issn = {2375-026X},
  doi = {10.1109/ICDE65448.2025.00198},
  urldate = {2026-01-23},
}

@article{zhao2023survey,
  title={A survey of large language models},
  author={Zhao, Wayne Xin and Zhou, Kun and Li, Junyi and Tang, Tianyi and Wang, Xiaolei and Hou, Yupeng and Min, Yingqian and Zhang, Beichen and Zhang, Junjie and Dong, Zican and others},
  journal={arXiv preprint arXiv:2303.18223},
  volume={1},
  number={2},
  year={2023}
}

@article{zhou2025comet,
author = {Zhou, Jiawei and Wang, Kai and Wang, Jianwei and Zhang, Kunpeng and Lin, Xuemin},
title = {COMET: An Interactive Framework for Efficient and Effective Community Search via Active Learning},
journal = {INFORMS Journal on Computing},
volume = {0},
number = {0},
pages = {null},
year = {0},
doi = {10.1287/ijoc.2024.0834},

URL = { 
    
        https://doi.org/10.1287/ijoc.2024.0834
    
    

},
eprint = { 
    
        https://doi.org/10.1287/ijoc.2024.0834
    
    

}
}

@article{fang2020survey,
  title={A survey of community search over big graphs},
  author={Fang, Yixiang and Huang, Xin and Qin, Lu and Zhang, Ying and Zhang, Wenjie and Cheng, Reynold and Lin, Xuemin},
  journal={The VLDB Journal},
  volume={29},
  pages={353--392},
  year={2020},
  publisher={Springer}
}

\end{document}